\documentclass[journal=jacsat,manuscript=article]{achemso}

\usepackage[version=3]{mhchem} 
\usepackage{graphicx}
\usepackage{amsmath}
\usepackage{amssymb}
\usepackage{color}
\usepackage[normalem]{ulem}
\usepackage{chemformula}
\usepackage{float}
\usepackage[symbol]{footmisc}
\usepackage{epstopdf}
\usepackage{physics}
\usepackage{bm}
\usepackage{hyperref}
\hypersetup{colorlinks=true,linkcolor=blue,citecolor=blue,urlcolor=blue}
\usepackage{booktabs}
\usepackage{textgreek}

\SectionNumbersOn

\DeclareGraphicsExtensions{.pdf,.png}

\author{Jing Shen}
\altaffiliation{These authors contributed equally to this work.}
\affiliation{Department of Chemistry, Shanghai Key Laboratory of Molecular Catalysis and Innovative Materials, State Key Laboratory of Porous Materials for Separation and Conversion, Fudan University, Shanghai 200438, P. R. China}

\author{Ziyan Ye}
\altaffiliation{These authors contributed equally to this work.}
\affiliation{Department of Chemistry, Shanghai Key Laboratory of Molecular Catalysis and Innovative Materials, State Key Laboratory of Porous Materials for Separation and Conversion, Fudan University, Shanghai 200438, P. R. China}

\author{Ming-Zheng Du}
\affiliation{Department of Chemistry, Shanghai Key Laboratory of Molecular Catalysis and Innovative Materials, State Key Laboratory of Porous Materials for Separation and Conversion, Fudan University, Shanghai 200438, P. R. China}

\author{Shi-Yu He}
\affiliation{Department of Chemistry, Shanghai Key Laboratory of Molecular Catalysis and Innovative Materials, State Key Laboratory of Porous Materials for Separation and Conversion, Fudan University, Shanghai 200438, P. R. China}

\author{Dong H. Zhang}
\affiliation{State Key Laboratory of Chemical Reaction Dynamics, Dalian Institute of Chemical Physics, Chinese Academy of Sciences, Dalian 116023, P. R. China}
\alsoaffiliation{University of Chinese Academy of Sciences, Beijing 100049, China}

\author{Jia-Xi Zeng}
\affiliation{
Dipartimento di Chimica, Universit\`{a} degli Studi di Milano, via Golgi 19, 20133 Milano, Italy}
\email{jiaxi.zeng@unimi.it}

\author{Venkat Kapil}
\affiliation{Department of Physics and Astronomy, London Centre for Nanotechnology, Thomas Young Centre, University College London, WC1E 6BT London, United Kingdom}
\email{v.kapil@ucl.ac.uk}

\author{Wei Fang}
\affiliation{Department of Chemistry, Shanghai Key Laboratory of Molecular Catalysis and Innovative Materials, State Key Laboratory of Porous Materials for Separation and Conversion, Fudan University, Shanghai 200438, P. R. China}
\email{wei_fang@fudan.edu.cn}

\title{
Rigorous Quantum Thermodynamics from Entropic Path Integral Coarse-Graining
}

\begin{document}






\newpage
\begin{abstract}
Nuclear quantum effects (NQEs) remain a major challenge for molecular simulations, as rigorous treatment requires imaginary-time path-integral methods with heavy computational overhead. 
Neglecting NQEs leads to systematic errors in thermodynamic properties and failures in predicting isotope effects, quantum tunnelling, and anharmonic zero-point motion.
Here, we introduce entropic path-integral coarse-graining (EPIGS), which enables rigorous quantum thermodynamics at the cost of classical simulations by training size- and temperature-transferable effective potentials utilising absolute centroid free energy and entropy.
Central to EPIGS is an instanton-based free-energy perturbation scheme that enables efficient and accurate evaluation of the centroid free energy and entropy for large systems, making construction of the EPIGS training dataset practical.
Benchmarks against full path-integral simulations on representative hydrogen-bonded systems, including liquid water, show that EPIGS reproduces quantum free energies and enthalpies within 0.2 meV/atom at near-classical computational cost. 
EPIGS provides a highly accurate, scalable and low-cost framework for quantum thermodynamic simulations of complex systems across temperatures.

\end{abstract}

\newpage
\section{Introduction}

Molecular dynamics (MD) is a cornerstone tool for obtaining atomistic insights into structure, dynamics, mechanisms, and thermodynamic properties.
Recent advances in machine-learning interatomic potentials (MLIPs)\cite{deringer_gaussian_2021, behler_four_2021, jacobs_practical_2025} have brought \textit{ab initio} quality accuracy to large-scale simulations at linear computational cost. 
Unfortunately, the overwhelming majority of these simulations treat the nuclei as classical particles, neglecting zero-point motion, tunnelling, and isotope effects.
As molecular simulations become increasingly accurate and predictive, the neglect of nuclear quantum effects (NQEs) is emerging as a major source of systematic error, compromising predictions of isotope and tunnelling effects in chemical reactions, phase behaviour of condensed matter, and relevant applications such as isotope separation, catalysis, and drug design.
Indeed, the crucial role of NQEs has been reported across diverse systems~\cite{Markland_NatRevChem_2018_v2_p109}, from various fundamental systems such as water and ice~\cite{Ceriotti_ChemRev_2016_v116_p7529,Benoit_Nature_1998_v392_p258, Chen_PhysRevLett_2003_v91_p215503, Morrone_PhysRevLett_2008_v101_p17801,Serra_PhysRevLett.108.193003, bore_realistic_2023}, interfacial and confined water \cite{Li_PhysRevLett_2010_v104_p66102,zhu_nanoconfined_ices_2024}, interfacial hydrogen \cite{Fang_IntRevPhysChem_2019_v38_p35}, gas-phase dimers \cite{Li_ProcNatlAcadSciUSA_2011_v108_p6369,Fang_JPhysChemLett_2016_v7_p2125}, molecular crystals~\cite{Rossi_PhysRevLett_2016_v117_p115702, Kapil_ProcNatlAcadSciUSA_2022_v119_pe2111769119}, and simple organic liquids~\cite{Pereyaslavets_ProcNatlAcadSciUSA_2018_v115_p8878,Ugur2025}, to more complex or application-oriented systems including biomolecules~\cite{masgrau_atomic_2006, perez_enol_2010, Wang_ProcNatlAcadSciUSA_2014_v111_p18454, Rossi2015,shen2025rolenuclearquantumeffects}, metal organic frameworks (MOFs) for hydrogen storage or isotope separation \cite{wahiduzzaman_hydrogen_2014, liu_barely_2019}, and superconducting systems \cite{errea_quantum_2016}.  \\

Imaginary-time path-integral methods are widely recognized for incorporating NQEs in atomistic simulations, notably through path-integral molecular dynamics (PIMD) and path-integral Monte Carlo (PIMC).
For quantum Boltzmann statistics, PIMD/PIMC exploits the isomorphism between the quantum partition function and that of a classical closed ring-polymer comprising multiple replicas (beads) of the physical system.~\cite{feynman_quantum_1965, chandler_exploiting_1981}.
Sampling a closed ring-polymer with many beads carries a substantial computational overhead, typically an order of magnitude higher than classical MD, and increases rapidly to several orders for light nuclei at low temperatures.
Considerable effort has been devoted to reducing the computational cost of PIMD, with strategies that reduce the required number of beads. 
These include coloured-noise thermostats that mimic quantum fluctuations \cite{PhysRevLett.103.030603,Ceriotti_PhysRevLett_2012_v109_p100604}, ring-polymer contraction \cite{markland_efficient_2008} often paired with multiple-time-step integrators~\cite{kapil_accurate_2016, marsalek_quantum_2017},
and higher-order splitting of the Boltzmann operator~\cite{jang_applications_2001, perez_improving_2011, kapil_high_2016}, or truncated cumulant expansions~\cite{takahashi_monte_1984, poltavsky_modeling_2016}. 
These methods have been effective and have considerably extended the practical applicability of PIMD~\cite{Markland_NatRevChem_2018_v2_p109}. 
However, their efficiency gains are typically system- and regime-dependent, and substantial reductions to near-classical cost often rely on assumptions about the potential energy surface and/or sacrifices in accuracy. 
The need to sample multiple replicas therefore remains, and eliminating the overhead associated with imaginary-time path integrals in a generally applicable manner while retaining rigorous quantum statistics remains an open challenge.

An alternative approach for incorporating NQEs is based on quantum effective potentials that modify the underlying potential energy surface, a concept that originates from the Feynman-Hibbs (FH) theory~\cite{feynman_quantum_1965} and perturbation theories introduced by Wigner~\cite{wigner_quantum_1932} and Kirkwood~\cite{kirkwood_quantum_1933}. 
Recent developments have revitalised this concept through machine learning.
Most notably, the path-integral coarse-graining (PIGS) approach leverages MLIPs to learn the effective force acting either on the centroid or on an individual bead, thereby replacing the full ring-polymer simulation with a classical MD simulation on the learned effective potential \cite{musil_quantum_2022,loose_centroid_2022,Wu2022}. 
However, previous PIGS effective potentials are typically trained only on centroid forces and lack information about the absolute centroid free energy, leaving free energy predictions inaccurate and relative free energies between different regions arbitrary. 
This is because obtaining absolute free energies is impractical, requiring at least an order of magnitude more PIMD sampling.
As a result, PIGS is not suitable for quantitatively computing quantum thermodynamic properties. 
This limitation also affects transferability across systems and system sizes, since jointly trained models predict arbitrary relative free energies between different configurations.  
An additional complication is that, unlike potential energy surfaces, centroid free-energy surfaces depend explicitly on temperature, which is neglected in PIGS. 
As a result, separate PIGS models need to be trained for each temperature, and the temperature derivatives required for thermodynamic estimators are not available.

In this work, we develop an instanton-based free-energy perturbation approach (RPI-FEP) for efficient and accurate estimation of the path-integral centroid free energy, thereby providing direct access to the quantum entropic contribution. 
Building on this, we introduce entropic path-integral coarse-graining (EPIGS), in which both centroid free energy and entropy are learned to construct size- and temperature-transferable centroid effective potentials.
The resulting framework yields rigorous and accurate quantum thermodynamic properties for systems including liquid water over a wide temperature range at the computational cost of classical MD, demonstrating EPIGS' potential to supplant PIMD across a broad range of applications.

\section{Theory}
\subsection{Centroid formulation of the quantum partition function}
\label{ss:centroid}
In the imaginary-time path integral formulation of quantum statistical mechanics, the partition function is isomorphic to that of a classical ring polymer (RP),
which can be reformulated in terms of the RP centroid, \cite{feynman_quantum_1965,voth_path-integral_1996}
\begin{equation}
\begin{split}
    Z&=\left(\frac{m}{2\pi\beta_P\hbar^2}\right)^{Pf/2} \int\text{d}\vb*{x}_1\cdots\int\text{d}\vb*{x}_P~\mathrm{e}^{-\beta_P U_\text{RP}(\vb*{x}_1,\dots,\vb*{x}_P)}\\
    &=\Lambda^{-f}\int\text{d}\vb*{x}_\text{c}\left[\frac{P^{Pf/2}}{\Lambda^{(P-1)f}}\int \text{d} \vb*{x}_1\cdots\int \text{d}\vb*{x}_P
    ~\delta_\text{c}\,\mathrm{e}^{-\beta_P U_\text{RP}}\right]\\
    &\equiv\Lambda^{-f}\int\text{d}\vb*{x}_\text{c}~\mathrm{e}^{-\beta A_\text{c}(\vb*{x}_\text{c};\beta)}
\end{split}
\end{equation}
in which $P$ is the number of beads, $\beta=1/k_\text{B}T$ is the inverse temperature and $\beta_P=\beta/P$, $f$ is the number of degrees of freedom, $\Lambda=\sqrt{2\pi\beta\hbar^2/m}$ is the thermal de Broglie wavelength, 
$U_\text{RP}=\sum_{i=1}^{P}V(\vb*{x}_i)+
\sum_{i=1}^{P}\sum_{j=1}^f\frac{m}{2\beta_P^2\hbar^2}(x_{i+1,j}-x_{i,j})^2$
(with cyclic boundary conditions $\vb*{x}_{P+1}\equiv\vb*{x}_{1}$) is the RP potential, $\delta_\text{c}\equiv\delta\left(\frac{1}{P}\sum_{i=1}^{P}\vb*{x}_i-\vb*{x}_\text{c}\right)$ constrains the RP centroid to $\vb*{x}_\text{c}$,
and $A_\text{c}$ is the centroid free energy, also known as the centroid potential of mean force.
For simplicity, we assume that all atoms have mass $m$; extension to unequal masses is straightforward using the mass-weighted coordinates. 
The partial derivative of $A_\text{c}$ with respect to the centroid coordinate gives the centroid force,
\cite{voth_path-integral_1996,tuckerman2023statistical,Wu2022}
\begin{equation}
    \vb*{f}_\text{c}(\vb*{x}_\text{c};\beta)= -\frac{\partial A_\text{c}}{\partial \vb*{x}_\text{c}}=-\left<\frac{1}{P}\sum_{i=1}^P\left.\frac{\partial V}{\partial \vb*{x}}\right|_{\vb*{x}_i}\right>_{U_\text{RP},\vb*{x}_\text{c}},
\end{equation}
where the bracket $\langle \cdots\rangle_{U_\text{RP},\vb*{x}_\text{c}}$ denotes the centroid-constrained ensemble average under the RP potential $U_\text{RP}$. 
In contrast, $A_\text{c}$ is not accessible in the same manner as $\vb*f_\text{c}$ and is significantly more challenging to obtain.
Since $\vb*{f}_\text{c}$ is accessible, performing MD using $\vb*{f}_\text{c}$ corresponds to centroid molecular dynamics (CMD)~\cite{Cao1994,cao_formulation_1994,voth_path-integral_1996,Shi2003}, a widely known approach for approximating quantum dynamics and reaction rates. 
Because CMD evolves a classical trajectory under an effective force, it closely resembles classical MD and therefore offers the prospect of incorporating NQEs at near-classical computational cost. 
A significant advance in this direction is the machine learning of centroid forces, obtained either from centroid-constrained PIMD evaluations \cite{Wu2022} or via force matching to instantaneous centroid forces sampled in PIMD \cite{loose_centroid_2022,musil_quantum_2022}, as realized in the PIGS framework \cite{musil_quantum_2022}, enabling efficient CMD simulations. 
However, without knowledge of the full centroid free energy $A_\text{c}$, these force-only approaches cannot rigorously recover quantum thermodynamic quantities, including internal energy, enthalpy, and free energy.
In addition, previous methods require training a separate potential for each temperature, limiting transferability and increasing computational overhead for simulations across different temperatures.
EPIGS is designed to overcome these limitations, which is enabled by a key development that achieves efficient and rigorous evaluation of $A_\text{c}$.

\subsection{
Instanton based free-energy perturbation
\label{ss:instanton_fep}
}
One standard route to obtain $A_\text{c}$ is through thermodynamic integration (TI) from classical to quantum nuclei, for example by mass-TI~\cite{Rossi2015,Fang_JPhysChemLett_2016_v7_p2125} or scaled-coordinates-TI~\cite{Habershon_TI_2011}. 
However, such TI-based approaches are generally too computationally demanding for this purpose, as obtaining $A_\text{c}$ typically requires around an order of magnitude more PIMD sampling than evaluating $\vb*{f}_\text{c}$.
Free-energy perturbation (FEP)~\cite{zwanzig_hightemperature_1954} provides an alternative route, computing the free energy difference between the true potential $U_b$ and a reference potential $U_a$ in one simulation using
\begin{equation}
A_{\text{c},b}-A_{\text{c},a}
=-\frac{1}{\beta}\ln\left\langle \mathrm{e}^{-\beta_P\left(U_b-U_a\right)}\right\rangle_{U_a,\vb*{x}_\text{c}}.
\label{FEP}
\end{equation}
The primary challenge in reliable FEP is constructing a reference potential $U_a$ with analytically known free energy and sufficient overlap with the configuration-space distribution of the true potential.
%

Ring-polymer instantons are periodic imaginary-time paths that minimise the Euclidean action, providing a semiclassical description of quantum tunnelling.
Inspired by semiclassical instanton theory \cite{RPInst,Andersson-2009-JPCA-4468,Kastner-2014-WIRE,InstReview}, we construct a ring-polymer-instanton (RPI)-like reference for FEP by identifying the minimal-action imaginary time path $\bm{\mathsf{x}}^*$ with its centroid constrained at $\vb*{x}_\text{c}$.
When the centroid configuration exhibits imaginary modes satisfying $\text{Im}\,\omega_j > \frac{2\pi}{\beta\hbar}$, the ring-polymer configuration delocalises along these unstable directions, mirroring the delocalised structure of instantons below the crossover temperature.
In analogy to instanton theory, the FEP reference potential is obtained by truncating the ring-polymer potential to second order about $\bm{\mathsf{x}}^*$,
\begin{equation}
    U_\text{RPI}(\bm{\mathsf{x}})=\left.U_\text{RP}\right|_{\bm{\mathsf{x}}^*}
    +\frac{1}{2}(\bm{\mathsf{x}}-\bm{\mathsf{x}}^*)^\mathrm{T} \left.\widetilde{\textbf{H}}_\text{RP}\right|_{\bm{\mathsf{x}}^*}(\bm{\mathsf{x}}-\bm{\mathsf{x}}^*),
\end{equation}
where $\bm{\mathsf{x}}\equiv(\vb*{x}_1,\dots,\vb*{x}_P)$ denotes the closed imaginary-time path subject to $\delta_\text{c}$, and $\widetilde{\textbf{H}}_\text{RP}$ is the ring-polymer Hessian with the centroid normal modes projected out.
We note that the centroid constraint $\delta_\text{c}$ strongly restricts the configuration space, making this local harmonic form a natural and effective reference for FEP.
Similar to instanton theory calculations, $\bm{\mathsf{x}}^*$ can be obtained using instanton optimisation algorithms \cite{richardson_2012} to minimize $U_\text{RP}$ with the constraint $\delta_\text{c}$.
The free energy of the RPI reference can be computed analytically,
\begin{equation}
    A_\text{c,RPI}(\vb*{x}_\text{c};\beta)=-\frac{1}{\beta}\ln \left[P^{f+1}\sqrt{\frac{ B_P}{2\pi\beta_P\hbar^2}}\prod_{k=2}^{(P-1)f}\frac{1}{\beta_P\hbar\eta_k}\right]+\frac{\left.U_\text{RP}\right|_{\bm{\mathsf{x}}^*}}{P}
    \label{AcRPI}
\end{equation}
in which $B_P=\sum_{i=1}^P\sum_{j=1}^f m_j(x^*_{i+1,j}-x^*_{i,j})^2$, $\bm{\mathsf{e}}_k$ and $\eta_k$ are the eigenvectors and eigenvalues of the mass-weighted $\left.\widetilde{\textbf{H}}_\text{RP}\right|_{\bm{\mathsf{x}}^*}$.
The product starts from $k=2$ since $\widetilde{\textbf{H}}_\text{RP}$ is positive semi-definite, with a single zero mode $\bm{\mathsf{e}}_1$ associated with permutational invariance, which is treated separately.
A coordinate transformation from $\boldsymbol{x}_i$ in the Cartesian basis to $(\tau_0,q_k)_{k\ge2}$ in $\{ \bm{\mathsf{e}}_k(\tau_0)\}_{k\ge2}$ basis, where $\tau_0\in[0,\beta\hbar)$ is a permutational coordinate, allows one to integrate out all possible permutations.
Consequently, sampling can be restricted to the eigenvector subspace with positive eigenvalues $\{\bm{\mathsf{e}}_k\}_{k\ge2}$, allowing FEP to be performed using path integral Monte Carlo (PIMC) with the Gaussian sampling protocol,
\begin{equation}
    A_\text{c}=A_\text{c,RPI}-\frac1\beta\ln\frac{\int\text{d}q_2\cdots\text{d}q_{(P-1)f}\,w\,\mathrm{e}^{-\beta_P(U_\text{RP}-U_\text{RPI})}\,\mathrm{e}^{-\beta_P U_\text{RPI}}}{\int\text{d}q_2\cdots\text{d}q_{(P-1)f}\,\mathrm{e}^{-\beta_P U_\text{RPI}}}.
\end{equation}
Additionally, a weight term $w$ arising from the Jacobian determinant of the coordinate transformation needs to be included (see the Supplementary Information (SI) Section S1 for the derivation).
We name this method the ring-polymer instanton free-energy perturbation (RPI-FEP) method.
For centroid geometries satisfying $\text{Im}\,\omega_j < \frac{2\pi}{\beta\hbar}$, the ``RPI" collapses, and $A_\text{c,RPI}$ returns to the $P$-bead RP free energy for the harmonic oscillator (HO) \cite{kleinert2009path} with the centroid constrained at $\vb*x_c$, 
\begin{equation}
A_\text{c,HO}\left(\vb*{x}_\text{c};\beta\right)=\frac1\beta\sum_{j=1}^{f}
\ln\frac{\sinh\left[P\sinh^{-1}\left(\beta_P\hbar\omega_j/2\right)\right]}{\beta\hbar\omega_j/2}+V(\vb*{x}_\text{c}).
\label{AcHO}
\end{equation}

\subsection{
Entropic path integral coarse graining
\label{ss:temperature_encoding}
}

\begin{figure}[tbp]
    \centering
    \includegraphics[width=0.7\textwidth]{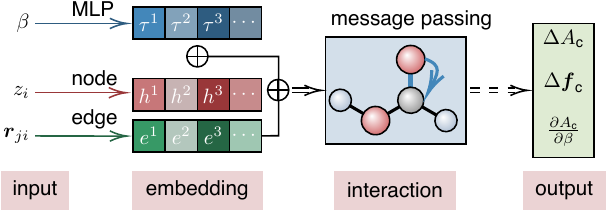}
    \caption{
    Schematic of temperature-dependent message-passing neural network architecture. 
    The inverse temperature $\beta$ is encoded via a multilayer perceptron (MLP) into a latent embedding vector $\boldsymbol{\tau}$, which is broadcast to all atoms and concatenated ($\oplus$) with the node features $\boldsymbol{h}_{i}$ (initialized from atomic numbers $z_i$) prior to message passing.
    Edge features $\boldsymbol{e}_{ji}$ encode interatomic distances and directions ($\boldsymbol{r}_{ji}$). 
    The network outputs the quantum contributions $\Delta A_\text{c}$ and $\Delta\boldsymbol{f}_\text{c}$, and the temperature derivative $\frac{\partial A_\text{c}}{\partial \beta}$.
    }
    \label{fig:temp_embed}
\end{figure}

The centroid entropy $S_\text{c}$ is the temperature derivative of $A_\text{c}$; therefore it serves as the key quantity for learning the temperature dependence of $A_\text{c}$.
With $A_\text{c}$ available, we can compute $S_\text{c}$ from the same centroid-constrained PIMD simulation from which we obtain $\vb*{f}_\text{c}$:
\begin{equation}
TS_\text{c}=\beta\frac{\partial A_\text{c}}{\partial \beta}=\left<\widehat{K}+\widehat{V}\right>_{U_\text{RP},\vb*{x}_\text{c}}-\frac{f}{2\beta}-A_\text{c},
\label{dAcdb}
\end{equation}
where $\widehat{K}$ and $\widehat{V}$ are the (virial) kinetic and potential energy estimators, respectively.
Utilising the centroid free energy and entropy, EPIGS leverages machine learning to construct size- and temperature-dependent centroid free energy surfaces in order to achieve rigorous simulation of quantum thermodynamics at the cost of classical MD.
In the following, we describe the three components of the EPIGS framework: dataset generation, model training, and inference.
\begin{enumerate}
    \item \textbf{Dataset generation} : For every given configuration $\vb{x}_{\text{c}}$ in the training set, a PIMC simulation and a centroid-constrained PIMD simulation are performed at a temperature $T$. 
    From these simulations, we obtain (i) the absolute free energy $A_c$ using the instanton FEP technique introduced in Section \ref{ss:instanton_fep} of this work, (ii) the centroid mean force which is the position derivative of $A_c$
    and  (iii) the entropic contribution to the free energy $T S_c$ which determines the temperature derivative of $A_c$.
    Furthermore, as done by ~\citet{musil_quantum_2022}, we apply a $\Delta$-learning strategy~\cite{ramakrishnan_big_2015}, training on the difference between the centroid free energy and the potential energy, $\Delta A_\text{c}(\boldsymbol{x}_\text{c};\beta)    \equiv A_\text{c}(\boldsymbol{x}_\text{c};\beta)-V(\boldsymbol{x}_\text{c})$,
    and the corresponding force differences, $\Delta \vb*{f}_\text{c}(\boldsymbol{x}_\text{c};\beta)\equiv \vb*{f}_\text{c}(\boldsymbol{x}_\text{c};\beta) - \vb*{f}(\boldsymbol{x}_\text{c})$. 
    $\Delta A_\text{c}$ and $\Delta \vb*{f}_\text{c}$ vary smoothly and spans a substantially smaller range than the full quantities, making it possible to achieve high accuracy with limited training data.
    A dataset
    \(
    \mathcal{D}
    =
    \left\{
    \left(
    \vb*{x}^{(j)}_\mathrm{c},
    T^{(j)},
    \Delta A_\mathrm{c}^{(j)},
    \Delta\vb*{f}_\mathrm{c}^{(j)},
    S_\mathrm{c}^{(j)}
    \right)
    \right\}_{j=1}^{M}
    \)
    is thus collected, where the inputs are the first two elements, and the corresponding labels are the latter three elements.

    \item \textbf{Training} : With the training dataset collected, we learn a scalar-valued function $\Delta\hat{A}_\mathrm{c} :  \mathbb{R}^{f}\to \mathbb{R}$ which establishes the mapping $(\vb{x}_\mathrm{c},T)\mapsto \Delta\hat{A}_\mathrm{c}(\vb{x}_\mathrm{c};T)$. 
    The centroid mean force correction and entropy are obtained as derivatives of $\Delta\hat{A}_\mathrm{c}$. 
    Specifically, for each training sample $j$, the model is trained to satisfy
    \begin{align}
    \Delta\hat{A}_\mathrm{c}\!\left(\vb{x}_\mathrm{c}^{(j)};T^{(j)}\right)
    &\approx \Delta A_\mathrm{c}^{(j)} \label{eq:c1},\\
    -\nabla_{\vb{x}_\mathrm{c}} \Delta\hat{A}_\mathrm{c}\!\left(\vb{x}_\mathrm{c}^{(j)};T^{(j)}\right)
    &\approx \Delta\vb{f}_\mathrm{c}^{(j)}\label{eq:c2},\\
    -\partial_T \Delta\hat{A}_\mathrm{c}\!\left(\vb{x}_\mathrm{c}^{(j)};T^{(j)}\right)
    &\approx S_\mathrm{c}^{(j)} \label{eq:c3}.
    \end{align}
    Entropy enforces thermodynamic consistency by coupling the learned free-energy surface to its temperature derivative, ensuring internally consistent predictions of both the centroid free energy and the quantum entropic term within a single training protocol.
    Standard MLIPs naturally enforce the energy and force consistency conditions Eq.~\ref{eq:c1} and Eq.~\ref{eq:c2}, but not the entropy condition Eq.~\ref{eq:c3}. 
    When trained at a single temperature, EPIGS reduces to a standard MLIP trained on centroid energies and forces.

    We modify the MACE equivariant message-passing neural network \cite{NEURIPS2022_4a36c3c5} to implement these functionalities, although EPIGS is readily extensible to other MLIP architectures.
    To enable temperature transferability, the EPIGS model incorporates $\beta$ as a continuous input via graph-level conditioning.
    A multilayer perceptron encodes $\beta$ into a latent vector that is broadcast to all nodes prior to message passing, preserving the intensive nature of temperature (Fig.~\ref{fig:temp_embed}).
    This construction allows the temperature derivative $\frac{\partial A_\text{c}}{\partial\beta}$ (essentially the quantum entropic term) to be obtained directly via automatic differentiation.
    The total training loss function combines energy, force, and entropic matching terms ($\mathcal{L}_A$, $\mathcal{L}_f$, and $\mathcal{L}_S$), $\mathcal{L} = w_A \mathcal{L}_A + w_f \mathcal{L}_f + w_{S} \mathcal{L}_S,$
    where the weights ($w_A$, $w_f$ and $w_S$) balance their relative contributions.\\

    \item \textbf{Inference.} $\Delta A_\text{c}$ is obtained from a forward pass, whereas $\Delta \vb*{f}_\text{c}$ and the entropic term $\frac{\partial A_\text{c}}{\partial \beta}$ are evaluated via automatic differentiation in a backward pass. 
    These quantities are used together with a base potential energy surface, with the EPIGS model providing corrections that transform the classical potential energy and forces into the centroid free energy and centroid forces.
    The resulting centroid mean force then drives the molecular dynamics, while the free energy and entropy are accumulated for post-processing.
    Without explicitly sampling the ring polymer, the quantum-mechanical internal energy, $U^\text{q}$, can be computed directly from an EPIGS-MD simulation as
    \begin{equation}
        U^\text{q}=-\frac{\partial\ln Z}{\partial \beta}
        =
        \frac{\int\text{d}\vb*{x}_\text{c}\,
        \left(A_\text{c}+\beta\frac{\partial A_\text{c}}{\partial \beta}\right)
        \mathrm{e}^{-\beta A_\text{c}(\vb*{x}_\text{c};\beta)}}
        {\int\text{d}\vb*{x}_\text{c}\,
        \mathrm{e}^{-\beta A_\text{c}(\vb*{x}_\text{c};\beta)}}
        +\frac{f}{2\beta}.
        \label{eq:U}
    \end{equation}
    Similarly, the classical-to-quantum free energy difference is given by
    \begin{equation}
        A^\text{q}-A^\text{cl}\equiv \Delta A^{\mathrm{q}\leftarrow \mathrm{cl}}
        =
        -\frac{1}{\beta}
        \ln
        \frac{\int\text{d}\vb*{x}_\text{c}\,
        \mathrm{e}^{-\beta A_\text{c}(\vb*{x}_\text{c};\beta)}}
        {\int\text{d}\vb*{x}_\text{c}\,
        \mathrm{e}^{-\beta V(\vb*{x}_\text{c})}},
    \end{equation}
    which can be computed via thermodynamic integration (TI) over an order parameter $\lambda\in[0,1]$ that smoothly transforms $V$ into $A_\text{c}$ through $V_\lambda = V + \lambda\left(A_\text{c}-V\right)$,
    a procedure we denote as EPIGS-TI. 
    The remaining thermodynamic potentials, namely the enthalpy $H^\text{q}$ and Gibbs free energy $G^\text{q}$, can be computed analogously in the isothermal-isobaric ensemble or using thermodynamic relations.
\end{enumerate}

\section{Results}
We test the EPIGS framework on several key thermodynamic properties of a set of representative systems of  increasing complexity: (i) the formic acid series, including formic acid dimer (FAD), deuterated formic acid HCOOD, and formate anion HCOO$^-$, (ii) the adenine–thymine (AT) base pair, and (iii) liquid water, chosen as fundamental hydrogen-bonded systems covering gas-phase clusters, biologically relevant motifs, and the condensed phase.
All simulations employ the MACE-OFF23 pretrained model \cite{Kovacs_JAmChemSoc_2025_v147_p17598} as the underlying potential, 
which is further fine-tuned via continued optimization~\cite{kaur_data-efficient_2024} for each system with additional data calculated at the {\textomega}B97M-D3(BJ)/def2-TZVPPD \cite{Najibi_JChemTheoryComput_2018_v14_p5725,Weigend_PhysChemChemPhys_2005_v7_p3297} level using the ORCA program\cite{ORCA6} (see SI Section S2.A for details).
We first assess the accuracy and efficiency of the RPI-FEP approach for computing $A_\text{c}$ in these systems, as this provides the training data required for EPIGS. 
We then examine how well EPIGS learns the resulting centroid free-energy surfaces and centroid entropy. 
Finally, we examine its practical utility for efficient and accurate quantum thermodynamic prediction by evaluating classical-to-quantum free-energy shifts $\Delta A^{\mathrm{q}\leftarrow \mathrm{cl}}$, gas-phase binding and dissociation free energies at room temperature, and the enthalpy of vaporization of liquid water over a wide temperature range. \\
%

\begin{figure}[tbp]
    \centering
    \includegraphics[width=1.0\textwidth]{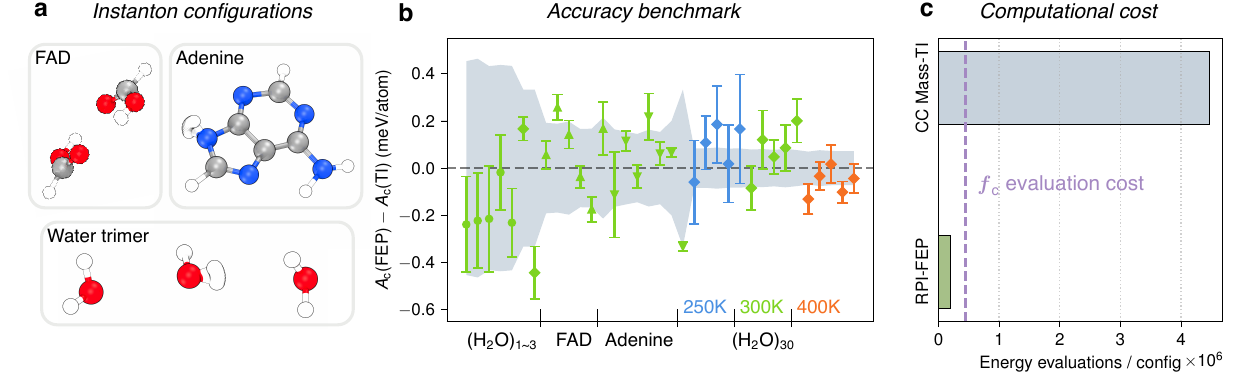}
    \caption{
    Validation and computational cost of the RPI-FEP method for centroid free energy evaluation. 
    (a) Representative ring-polymer instanton configurations for formic acid dimer (FAD), adenine, and water trimer, illustrating nuclear quantum delocalisation of hydrogen atoms.
    (b) Benchmark of RPI-FEP against centroid-constrained mass-TI (CC Mass-TI) for the centroid free energy.
    Deviation $A_{\text{c}}(\text{FEP}) -A_{\text{c}}(\text{TI})$ is shown for molecules and small clusters ((H$_{2}$O)$_{1\sim 3}$, FAD, adenine) at 300 K, and for large 30-molecule water clusters at 250, 300, and 400 K,
    with marker colours indicating temperatures.
    Marker shapes distinguish systems: circles (water monomer), diamonds (water clusters), upward triangles (FAD), and downward triangles (adenine).
    Systems are ordered by increasing sizes, ranging from 3 to 90 atoms.
    Shaded region indicates the statistical uncertainty of the mass-TI reference.
    (c) Computational cost per configuration. The cost is measured as effective number of potential energy evaluations, with an overhead factor of 1.5 for energy-and-force evaluations relative to energy-only evaluations, reflecting the additional cost of the backward pass for gradient computation.
    }
    \label{fig:fep_benchmark}
\end{figure}

We first benchmark RPI-FEP for computing absolute free energies of frozen centroid configurations against centroid-constrained mass-TI across the test systems and temperatures, before assessing the performance of EPIGS.
Simulation details and convergence tests are presented in SI Sections S2.C-D  and S3.A-B, respectively. 
The test configurations are randomly selected from FAD, adenine, and water clusters of varying sizes, spanning diverse hydrogen-bonding environments and NQEs.
Among them, five configurations exhibit large imaginary frequencies, with delocalised RPI reference states.
These instanton-like references all exhibit delocalisation of a single hydrogen atom along a dominant imaginary mode (see e.g. Fig.~\ref{fig:fep_benchmark}a). 
As shown in Fig.~\ref{fig:fep_benchmark}b, the RPI-FEP estimates of the centroid potential $A_\text{c}$ are in excellent agreement with the benchmark results across all test configurations, with nearly all RPI-FEP values falling within the statistical error bars of mass-TI. 
Importantly, this accuracy is achieved at a dramatically reduced computational cost: 
while centroid-constrained mass-TI is approximately an order of magnitude more expensive than the calculation of $\boldsymbol{f}_\text{c}$ by a centroid-constrained PIMD, computing $A_\text{c}$ via RPI-FEP costs only $\sim$44\% as much as computing $\boldsymbol{f}_\text{c}$ (see Fig.~\ref{fig:fep_benchmark}c).
These results demonstrate that the RPI-FEP approach provides a reliable and highly efficient estimate of the key quantity $A_\text{c}$ for EPIGS.

Having established the accuracy and efficiency of our approach for computing $A_c$, we next examine the performance of the EPIGS models trained at a single temperature for the formic acid and AT base pair systems.
Training data were generated from PIMD trajectories by selecting 150–300 representative configurations for each species using farthest-point sampling to maximise structural diversity (SI Section S2.B).
No energy-based filtering was applied, so the dataset naturally spans the full range of thermally accessible configurations.
For each cluster, the corresponding monomer(s) are trained together in a single EPIGS model. 
The EPIGS models achieve excellent training accuracy and size-transferability across all systems (see SI Section S4 for parity plots and full error tables), with root-mean-square errors (RMSEs) below 0.1 meV per atom for $\Delta A_\text{c}$ (except for HCOOD with a RMSE of 0.14 meV per atom) and small relative force errors of 4-7\% across all systems.
The consistently low training errors confirm energy–force consistency in the training data.
Validation and test errors closely match training performance for both $\Delta A_\text{c}$ and $\Delta \boldsymbol{f}_\text{c}$, indicating minimal overfitting and robust generalisation. 
The uniform accuracy observed across chemically and structurally diverse systems highlights the broad applicability of the EPIGS framework. \\

\begin{figure}[tbp]
    \centering
    \includegraphics[width=1.0\textwidth]{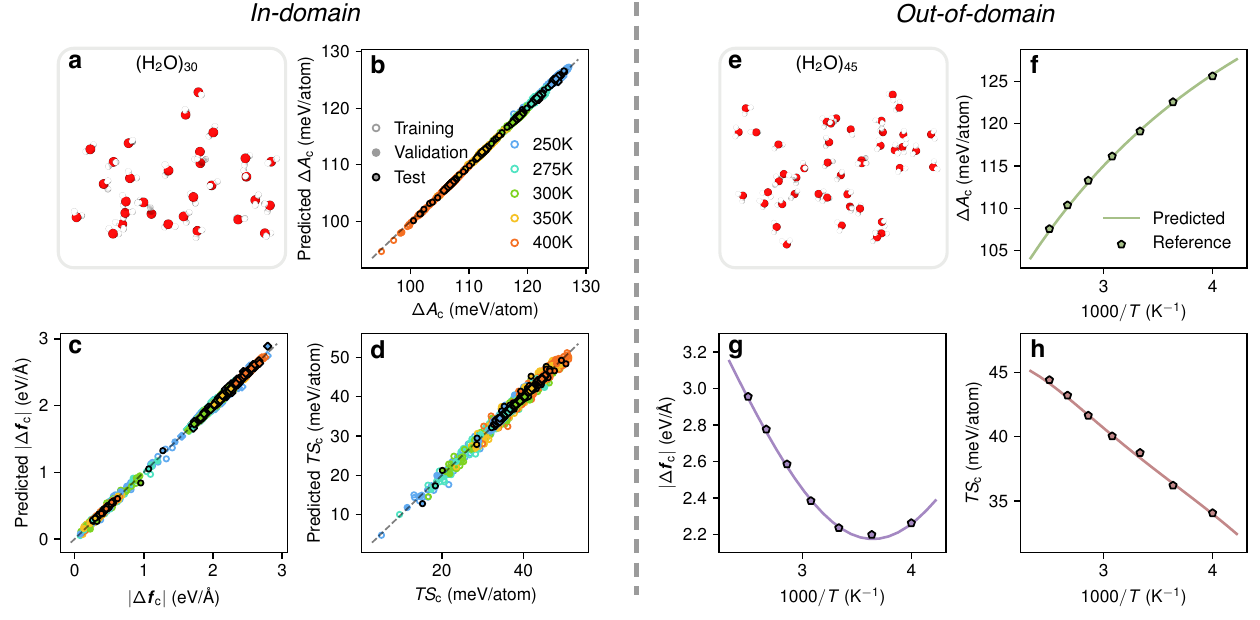}
    \caption{
    Accuracy and transferability of the EPIGS model for water. 
    (a) Representative snapshot of a 30-molecule water cluster from the training set. 
    (b–d) Parity plots of EPIGS model predictions against reference values across five temperatures (250--400 K, colour-coded) for (b) the centroid free energy shift $\Delta A_{\text{c}}$, (c) the centroid force magnitude $| \Delta \boldsymbol{f}_{\text{c}}| $, and (d) the entropic term $TS_{\text{c}}$.
    Marker shapes distinguish monomer (circles) and cluster (diamonds) configurations.
    (e) Snapshot of a larger 45-molecule water cluster not included in training.
    (f-h) EPIGS predictions of (f) $\Delta A_{\text{c}}$, (g) $| \Delta \boldsymbol{f}_{\text{c}}| $, and (h) $TS_{\text{c}}$ for the 45-molecule test cluster as a function of inverse temperature (solid line) v.s. reference values obtained from RPI-FEP and centroid-constrained PIMD calculations (markers).
    }
    \label{fig:water_parity}
\end{figure}

After establishing performance in simpler systems, we advance to the construction of size- and temperature-transferable EPIGS for liquid water, which is the most challenging among the systems considered.
Training data comprise both water monomers and large clusters containing 30 molecules (Fig.~\ref{fig:water_parity}a), with dataset size, partitioning, and sampling protocol specified in SI Section S2.B.
Centroid configurations were generated at 300 K, with $\Delta A_\text{c}$ and $\Delta \vb*f_\text{c}$ computed at five temperatures (400, 350, 300, 275, and 250 K) spanning above the boiling point and below the freezing point of water.
As shown in Fig.~\ref{fig:water_parity}b and c, EPIGS accurately learns $\Delta A_\text{c}$ and $\Delta \boldsymbol{f}_\text{c}$ across all temperatures, achieving RMSEs in $\Delta A_\text{c}$ of 0.1–0.2 meV per atom for the intermediate temperatures and $\sim$0.3 meV per atom at the temperature extremes (Table S3).
Similarly strong performance is observed for the quantum entropic term $TS_\text{c}$ (Fig.~\ref{fig:water_parity}d), with RMSEs below 0.3 meV/atom for the cluster and around 1 meV/atom for the monomer.
This level of accuracy is notable given that the training does not include information on the second-derivative with respect to $\beta$. 
Again, validation and test errors closely follow training performance, indicating minimal overfitting. 
The parity plots also show no systematic deviation between prediction and reference.
More importantly, EPIGS remains predictive beyond the training domain. 
For an unseen 45-molecule water cluster (Fig.~\ref{fig:water_parity}e), it achieves quantitative agreement with reference calculations for $\Delta A_\text{c}$, $\Delta\boldsymbol{f}_\text{c}$, and $TS_\text{c}$ across the full temperature range (Fig.~\ref{fig:water_parity}f-h), indicating robust size-transferability of the learned quantum thermodynamic mapping to liquid water.

\begin{figure}[tbp]
    \centering
    \includegraphics[width=0.7\textwidth]{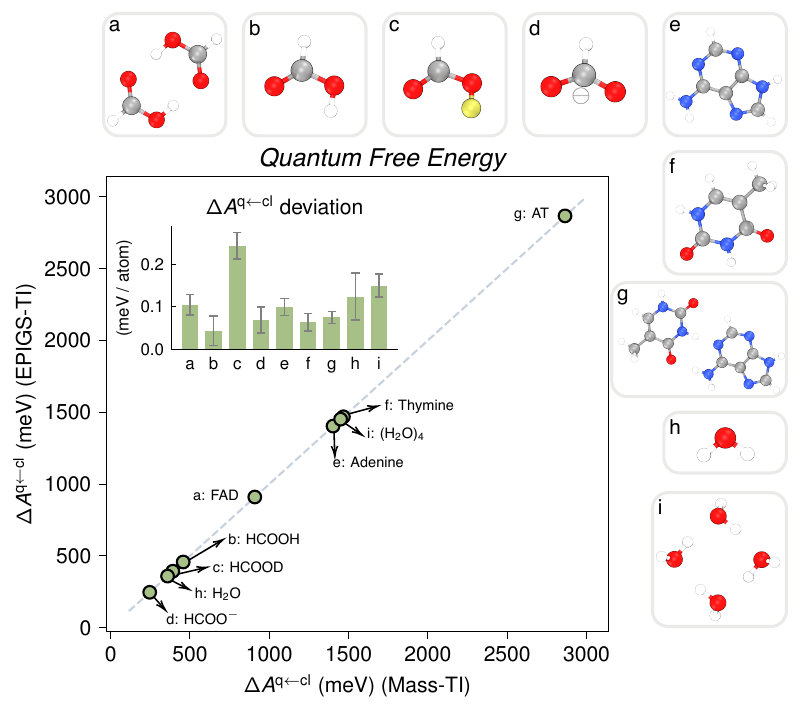}
    \caption{
    Application of EPIGS-TI to quantum free energies of gas-phase clusters.
    Panels (a–i): Snapshots for all benchmark systems
     (atom colours: C gray, N blue, O red, H white, D yellow).
    The central parity plot compares EPIGS-TI against mass-TI for the classical-to-quantum free energy changes $\Delta A^{\text{q}\leftarrow\text{c}}$ for all the systems.
    The inset shows the per-atom deviation between the two methods for each system.
    }
    \label{fig:free_energy_benchmark}
\end{figure}

With accurate and robust EPIGS models in hand, we now evaluate their performance for predicting quantum thermodynamic properties. 
We benchmark on nine systems: formic acid series (FAD, HCOOH, HCOOD, HCOO$^{-}$), the adenine–thymine system (adenine, thymine, and the AT base pair), and water (monomer and tetramer).
Simulation setups and convergence tests are provided in SI Section S2.F and S3.C.
Fig. \ref{fig:free_energy_benchmark} compares the classical-to-quantum free energy shifts $\Delta A^{\text{q}\leftarrow\text{cl}}$ computed via EPIGS-TI against the mass-TI benchmark. 
Across all systems, EPIGS-TI reproduces the mass-TI reference to within 0.25 meV/atom, with most systems having errors below 0.1 meV/atom, which is comparable to the training error of EPIGS itself. 
The relative errors are below 0.1\% for all systems except HCOOD, demonstrating essentially-exact recovery of the quantum mechanical free energy.

We next consider the highly nontrivial task of predicting NQE contributions to binding and dissociation free energies. 
These quantities arise from differences between the classical-to-quantum free-energy shifts of related systems.
We compute quantum shifts for the binding free energy of FAD, AT base pair, and the water tetramer, and to the acid dissociation free energy for HCOOH and HCOOD.
In particular, NQE contributions to binding are small (albeit important), typically on the order of 1–10 meV per H-bond, placing stringent demands on model accuracy.
As shown in Table~\ref{tab:delta_delta_A_G}, EPIGS-TI achieves surprisingly precise agreement within 1 meV of the mass-TI reference for all systems. 
Notably, the water tetramer binding energy is accurately reproduced despite the model being trained with no explicit tetramer data, demonstrating robust transferability across cluster sizes.
The results reveal that NQEs strengthen the binding of FAD and the AT base pair, while they destabilise the water tetramer.
For formic acid, NQEs lower the $\mathrm{p}K_\text{a}$ by 3.1 and yield an H$\rightarrow$D isotope shift of +1.1, underscoring the significant influence of nuclear quantum fluctuations on acid--base equilibria.
The ability of EPIGS to accurately capture delicate NQE contributions satisfies the demanding precision requirements of binding and dissociation thermodynamics, underscoring its broad application potential.

\begin{table}[tbp]
    \centering
    \caption{
    Nuclear quantum contributions to binding and dissociation free energies computed via EPIGS-TI and mass-TI.
    The binding free energy differences ($\Delta \Delta A_{\text{bind}}^{\text{q}\leftarrow \text{cl}}$) are shown for the molecular clusters (FAD, AT base pair, and water tetramer), while the acid dissociation free energy differences ($\Delta \Delta G_{\text{diss}}^{\text{q}\leftarrow \text{cl}}$) are shown for HCOOH and HCOOD, 
    with the $pV$ term evaluated under the ideal gas approximation.
    All values are in meV, with statistical uncertainties in parentheses applying to the last digit.
    }
    \begin{tabular}{lrrr}
    \toprule 
    Property  & Mass-TI & EPIGS-TI & Deviation \\
    \midrule\midrule 
     $\Delta \Delta A_{\text{bind} ,\ \text{FAD}}^{\text{q}\leftarrow \text{cl}}$ & $-5.8( 3)$ & $-5.16( 8)$ & $0.6( 3)$ \\
    $\Delta \Delta G_{\text{diss} ,\ \text{HCOOH}}^{\text{q}\leftarrow \text{cl}}$ & $-185.2( 2)$ & $-185.12( 5)$ & $0.1( 2)$ \\
    $\Delta \Delta G_{\text{diss} ,\ \text{HCOOD}}^{\text{q}\leftarrow \text{cl}}$ & $-119.9( 2)$ & $-120.87( 5)$ & $-0.9( 2)$ \\
    $\Delta \Delta A_{\text{bind} ,\ \text{AT}}^{\text{q}\leftarrow \text{cl}}$ & $-4.9( 6)$ & $-5.1( 1)$ & $-0.2( 6)$ \\
    $\Delta \Delta A_{\text{bind} ,\ {( \text{H}_{2} \text{O})_{4}}}^{\text{q}\leftarrow \text{cl}}$ & $+14.1( 5)$ & $+14.38( 5)$ & $0.3( 5)$ \\
     \bottomrule
    \end{tabular}
    \label{tab:delta_delta_A_G}
\end{table}

\begin{figure}[tbp]
    \centering
    \includegraphics[width=1.0\textwidth]{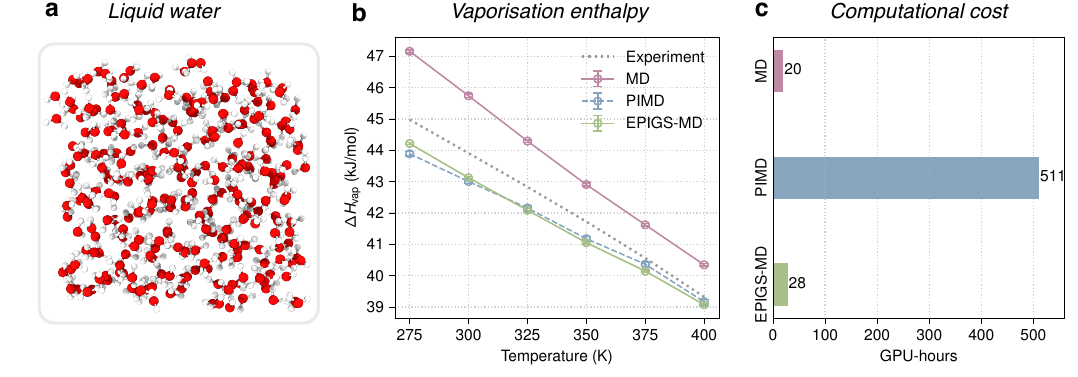}
    \caption{
    Application of EPIGS-MD to quantum enthalpy of liquid water.
    (a) Simulation snapshot of liquid water (270 molecules).
    (b) Enthalpy of vaporization $\Delta H_\text{vap}$ of liquid water as a function of temperature from classical MD (pink solid line), PIMD (blue dashed line), and EPIGS-MD (green solid line) simulations, compared with experimental values (grey dotted line) from the IAPWS-95 equation of state \cite{Wagner_JPhysChemRefData_2002_v31_p387}.
    (c) Computational cost (GPU-hours) for 1 ns simulation of liquid water.
    }
    \label{fig:water_enthalpy}
\end{figure}

Finally, we address the most challenging application of EPIGS: predicting the enthalpy of vaporization $\Delta H_{\text {vap }}$ for liquid water (Fig.~\ref{fig:water_enthalpy}a) across a temperature range. 
The challenge is three-fold: I. size-transferability: EPIGS is trained exclusively on clusters yet must predict bulk liquid properties; II. accurate $\Delta H_{\text {vap }}$ requires accurate predictions of not only $\Delta A_c$ but also the entropic term; and III. temperature-transferability: I and II must hold across the full temperature range.
At 300 K, classical MD predicts $\Delta H_{\text {vap }}$ to be $45.74(6) \mathrm{~kJ} / \mathrm{mol}$.
PIMD results show that NQEs raise the internal energy of the liquid more than that of the gas phase due to ZPE in intermolecular vibrational modes arising from the dense hydrogen-bond network, leading to a reduction in $\Delta H_{\text {vap }}$ by approximately $2.7 \mathrm{~kJ} / \mathrm{mol}$ relative to classical MD.
EPIGS-MD accurately reproduces the PIMD internal energies of both gas-phase and liquid water, yielding $\Delta H_{\text {vap }}=43.12(4) \mathrm{kJ} / \mathrm{mol}$ in excellent agreement with the PIMD value.
Remarkably, this quantitative agreement persists across the full temperature range from 275 to 400 K (Fig. \ref{fig:water_enthalpy}b), demonstrating reliable temperature-dependent predictions of EPIGS.
Meanwhile, EPIGS-MD requires only $\sim$40\% overhead relative to classical MD and remains more than an order of magnitude cheaper than PIMD (Fig. \ref{fig:water_enthalpy}c), thereby enabling rigorous quantum thermodynamic simulations at scales previously accessible only to classical MD.

Before concluding, we place EPIGS in the broader context of quantum simulations.
The main limitation of EPIGS is that it does not recover nuclear distributions associated with the path-integral replicas, and therefore cannot reproduce hydrogen positional distributions for which explicit PIMD remains necessary (although we note that it is possible to relate centroid distributions to bead distributions~\cite{cao_formulation_1994,sadhasivamEffectiveClassicalPotential2026}). 
Quantum kinetic and potential energy estimators also depend on bead configurations, thus cannot be obtained with the current EPIGS framework.
However, based on the relation\cite{Ceriotti_JChemPhys_2013_v138_p14112} 
\begin{equation}
    \frac{\langle \widehat{K}\rangle_{U_\text{RP},\vb*{x}_\text{c}}}{m}=-\frac{\partial A_\text{c}}{\partial m},
\end{equation}
it is possible to further extend EPIGS for these two properties.
EPIGS is particularly well suited for applications where the primary interest lies in rigorous quantum thermodynamic properties, approximate quantum dynamics via CMD, or quantum heavy-atom configurations, for which the centroid distribution closely approximates the bead distribution. 
In these cases, EPIGS provides a practical alternative to PIMD at the cost of classical MD.
Another key advantage is transferability across system sizes: models trained on small clusters or fragments can be deployed to predict condensed-phase or large-scale behaviour, enabling scalable simulations of complex systems. 
We also demonstrated reliable temperature-transferable learning of NQEs down to 250 K, with no fundamental limitations for applications at lower temperatures.
Therefore, the present work opens a path toward universal pre-trained EPIGS models spanning wide temperature ranges and chemical environments, providing a general and efficient route to incorporate NQEs in large-scale molecular simulations.

\section{Conclusions}
By explicitly incorporating centroid entropy, we introduce an entropic path-integral coarse-graining framework for rigorous quantum thermodynamics and size- and temperature-transferable learning of centroid potentials. 
This advance is made possible by the RPI-FEP approach, which provides an efficient and accurate calculation of the path-integral centroid free energy $A_\text{c}$, giving access to quantum entropy.
Across representative systems, EPIGS models reproduce the training data with sub-meV per-atom errors in $A_\text{c}$ and quantum entropy over the full temperature range considered. 
EPIGS yields precise quantum thermodynamic properties, including classical-to-quantum free energies within 0.2 meV per atom, quantum contributions to binding and dissociation free energies resolved with sub-meV accuracy, and an enthalpy of vaporization of liquid water in agreement with PIMD despite training exclusively on gas-phase clusters. 
Overall, EPIGS provides a rigorous and efficient framework for quantum thermodynamic simulations at near-classical computational cost, enabling practical large-scale simulations of NQEs and being capable of replacing PIMD in many thermodynamic applications.

\section*{Conflicts of interest}
There are no conflicts to declare.

\begin{acknowledgement}
This work is supported by the National Natural Science Foundation of China under grant number 22373021.
\end{acknowledgement}





\bibliography{ref}

\end{document}